\documentclass[sigconf]{acmart}

\renewcommand\footnotetextcopyrightpermission[1]{} 
\usepackage{booktabs} 
\usepackage{makecell}
\usepackage{url}
\usepackage{paralist}
\usepackage{fancyvrb}
\usepackage{amsmath}
\usepackage{epsfig}
\usepackage{subfig}
\usepackage{float}
\usepackage{tabularx}
\usepackage{amssymb}
\usepackage{graphicx}
\usepackage{color}

\newcommand{\q}[1]{\lq\lq{}{}#1\rq\rq{}{}}
\long\def\comment#1{}

\begin{document}

\title{Ranking Archived Documents for Structured Queries on Semantic Layers}

\author{Pavlos Fafalios}
\orcid{0000-0003-2788-526X}
\affiliation{
  \institution{L3S Research Center,\\Leibniz University of Hannover}
  \city{Hannover, Germany}}
\email{fafalios@L3S.de}

\author{Vaibhav Kasturia}
\affiliation{
  \institution{L3S Research Center,\\Leibniz University of Hannover}
  \city{Hannover, Germany}
}
\email{kasturia@L3S.de}

\author{Wolfgang Nejdl}
\affiliation{
  \institution{L3S Research Center,\\Leibniz University of Hannover}
  \city{Hannover, Germany}}
\email{nejdl@L3S.de}

\begin{abstract}
Archived collections of documents (like newspaper and web archives) serve as important information sources in a variety of disciplines, including Digital Humanities, Historical Science, and Journalism. However, the absence of efficient and meaningful exploration methods still remains a major hurdle in the way of turning them into usable sources of information. A semantic layer is an RDF graph that describes metadata and semantic information about a collection of archived documents, which in turn can be queried through a semantic query language (SPARQL). This allows running advanced queries by combining metadata of the documents (like publication date) and content-based semantic information (like entities mentioned in the documents). However, the results returned by such structured queries can be numerous and moreover they all equally match the query. 
In this paper, we deal with this problem and formalize the task of {\em \lq\lq{}{}ranking archived documents for structured queries on semantic layers\rq\rq{}{}}. Then, we propose two ranking models for the problem at hand which jointly consider: i) the relativeness of documents to entities, ii) the timeliness of documents, and iii) the temporal relations among the entities. The experimental results on a new evaluation dataset show the effectiveness of the proposed models and allow us to understand their limitations.
\end{abstract}



\maketitle

\section{Introduction}

Despite the increasing number of digital archives worldwide
(like news and web archives), the absence of efficient and meaningful
exploration methods still remains the major bottleneck in the way of
turning them into usable information sources \cite{calhoun2014exploring}.
Semantic models try to solve this problem by offering the means
to describe and publish metadata and semantic information
about a collection of archived documents in the standard RDF format.
A repository of such data, called Semantic Layer \cite{fafalios2017SemLayer},
allows running advanced queries which
combine metadata of the documents (like publication date) and
content-based semantic information (like entities mentioned in the documents).
For example, we can access a Semantic Layer over a newspaper archive and
find articles of a specific time period discussing about a specific category of entities
(e.g., {\em philanthropists}) or about entities that share some characteristics
(e.g., {\em lawyers born in Germany}), while we can also integrate information coming from other knowledge bases like DBpedia.
Such advanced information needs can be directly expressed through structured (SPARQL) queries
or through user-friendly interactive interfaces
which transparently transform user interactions to SPARQL queries
(e.g., Faceted Search-like browsing interfaces \cite{tzitzikas2016faceted}).

However, the results returned by such queries can be numerous and
moreover they all equally match the query: there is no relevance ranking like in the case of keyword-based information retrieval.
Thus, there arises the need for an effective method to rank the returned results
for discovering and showing to the users the most important ones.
For instance, when requesting articles from a news archive published within a specific time period and mentioning one or more query entities,
important documents may be those whose main topic is about an important event related to the query entities during the requested time period. 
Thus, an effective ranking method should
consider the different factors that affect the importance of documents to the query,
while at the same time relying only on the data
available in the semantic layer (there is no access to the full contents of the documents).

Although there is a plethora of works on ranking archived documents for keyword-based temporal queries, the problem of ranking such documents for the case of structured queries on knowledge graphs has not yet been recognized and studied. 
In this paper, we address this gap by first introducing and formalizing this type of problem.
Then, to cope with this problem, we propose two ranking models, a probabilistic one and a Random Walk-based one, which jointly consider the following aspects:
i) the {\em relativeness} of a document to the query entities,
ii) the {\em timeliness} of a document's publication date,
iii) the temporal {\em relatedness} of the query entities to other entities
mentioned in the documents.
The idea is to promote documents that
mention the query entities many times,
that have been published in important (for the query entities) time periods, and that
mention many other entities co-occurring frequently
with the query entities in important time periods.
Such an approach is widely applicable since it exploits the least amount of metadata.
 
In a nutshell, in this paper we make the following contributions:
\begin{itemize}
\item   We formulate and formalize the problem of ranking archived documents for structured queries over semantic layers.
\item   Due to lack of evaluation datasets for this problem, we have created a new ground truth dataset for a news archive which we make publicly available.
\item   We propose two ranking models for the problem at hand: a probabilistic one and a stochastic one (Random Walk with restart). 
\item   We present the results of an experimental evaluation which illustrate the effectiveness of the proposed models. We also analyze problematic cases for understanding when and why the models fail to provide good rankings. 
\end{itemize}

The rest of the paper is organized as follows:
Section \ref{sec:rw} presents the required background and related literature.
Section \ref{sec:problemDef} defines the problem and describes a new evaluation dataset.
Section \ref{sec:probabilisticModeling} introduces the probabilistic model. 
Section \ref{sec:stochasticModeling} introduces the stochastic model. 
Section \ref{sec:evaluation} presents evaluation results. 
Section \ref{sec:conclusion} concludes the paper and discusses interesting directions for future research.

\section{Background and Related Works}
\label{sec:rw}

\subsection{Semantic Layers}

\begin{figure*}
\vspace{-2mm}
\centering
\fbox{\includegraphics[width=6.8in]{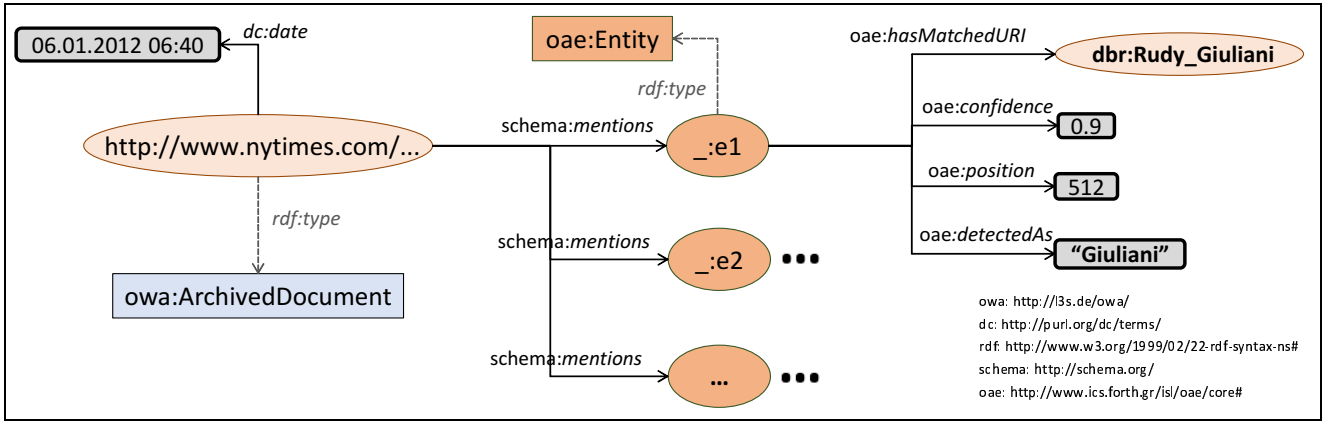}}
\vspace{-2mm}
\caption{A part of a Semantic Layer describing metadata and entity annotations for a news article.}
\label{fig:semanticLayerExample}
\end{figure*}

A Semantic Layer is an RDF repository (RDF graph) of structured data about a collection of archived documents \cite{fafalios2017SemLayer}. Structured data includes not only metadata information about a document (like publication date), but also {\em entity annotations}, i.e., disambiguated entities mentioned in each document extracted using an entity linking system \cite{shen2015entity}. 
Figure \ref{fig:semanticLayerExample} shows an example of (a part of) a Semantic Layer describing metadata and annotation information for a news article. We notice that the document was published on 6 January 2012 and mentions the entity name \q{Giuliani} at character position 512, which probably (with confidence score 0.9/1.0) corresponds to the known American lawyer and former politician Rudy Giuliani. 

A semantic layer allows running advanced, entity-centric queries that can also directly integrate information from other knowledge bases like DBpedia.
Listing \ref{listing:queryExampleIntegrate} shows an example of a SPARQL query that can be answered by a semantic layer over a collection of old news articles. The query requests articles of 1989 mentioning New York lawyers born in Brooklyn.
By accessing DBpedia at query-execution time, the query retrieves the entities that satisfy the query as well as additional information, in particular the birth date of each lawyer. We see that, using the entity URIs and type/category information,  we can find documents about entities even if we do not know the names of the entities, or without needing to specify a long list of all the entity names. Moreover, using the SPARQL operators {\tt COUNT} and {\tt GROUP BY}, and exploiting the expressive power of SPARQL, we can aggregate information related to documents and entities.

\renewcommand{\figurename}{Listing}
\setcounter{figure}{0}
\begin{figure}
\centering \scriptsize
\begin{Verbatim}[frame=lines,numbers=left,numbersep=1pt]
 SELECT ?article ?title ?date ?nylawyer ?bdate WHERE {
   SERVICE <http://dbpedia.org/sparql> {
     ?nylawyer dc:subject dbc:New_York_lawyers ; dbo:birthPlace dbr:Brooklyn .
     OPTIONAL { ?nylawyer dbo:birthDate ?bdate } }
   ?article dc:date ?date FILTER(year(?date) = 1989) .
   ?article schema:mentions ?entity .
   ?entity oae:hasMatchedURI ?nylawyer .
   ?article dc:title ?title
 } ORDER BY ?nylawyer
\end{Verbatim}
\vspace{-3mm}
\caption{Example SPARQL query over a Semantic Layer of a collection of old news articles. The query requests articles of 1989 discussing about New York lawyers born in Brooklyn.}
\label{listing:queryExampleIntegrate}
\end{figure}

As shown in \cite{fafalios2017SemLayer}, a semantic layer can answer information needs that existing keyword-based systems (like Google news) are not able to sufficiently satisfy. Such advanced (but also common) information needs can be directly expressed through SPARQL queries or by exploiting a user-friendly interface that transforms user interactions to SPARQL queries, like Sparklis \cite{ferre2014sparklis} or SemFacet \cite{arenas2014semfacet}.
Since the results returned by such structured queries can be numerous and have no ranking, in this paper we study how we can rank them based on their importance to the query entities.

\subsection{Related Works}
\label{rw}

We report works on the related problems of {\em time-aware document ranking} and {\em ranking in knowledge graphs}, and we discuss the commonalities and differences of these research areas to our problem.

\subsubsection{Time-aware Document Ranking}
\label{subsec:TimeDocumentRanking}

The impact of {\em  time} on information retrieval has received a large share of attention in the last decade. The surveys in \cite{campos2015survey} and \cite{kanhabua2015temporal} provide a comprehensive categorization and overview of temporal IR approaches and related applications. As regards time-aware {\em ranking}, existing works are classified into two different types based on two main notions of relevance with respect to time \cite{campos2015survey,kanhabua2015temporal}: 1) recency-based ranking, and 2) time-dependent ranking. Since recency-based ranking methods promote documents that are recently created or updated (this preference of freshness is common in general web searching), below we discuss only time-dependent ranking methods which are useful when searching archived collections of documents.

Jin et al. \cite{jin2008tise} proposed a ranking algorithm to sort results by applying a linear interpolation of text similarity, temporal information, and page importance (based on PageRank), where temporal similarity is the ranking score of temporal relevance based on the set of intersection conditions between the temporal query and the temporal expressions found in the web page. Arikan et al. \cite{arikan2009time} and Berberich et al. \cite{berberich2010language} introduced approaches that integrate temporal expressions extracted from the documents into language modeling frameworks. 
Metzler et al. \cite{metzler2009improving} proposed a time-dependent ranking model to adjust the document scores based on an analysis of web query logs and a set of document fields to estimate the time of both the query and the document. The work by Perkio et al. \cite{cataldi2010emerging} automatically detects topical trends and their importance over time within a news corpus using a statistical topic model and a simple variant of TF-IDF. These trends are then used as the basis for temporally adaptive rankings. 
Dakka et al. \cite{dakka2012answering} consider the publication time of documents to identify the important time intervals that are likely to be of interest to an implicit temporal query. Then, time is incorporated into language models to assign an estimated relevance value to each time period.
Aji et al. \cite{aji2010using} proposed a term weighting model that uses the revision history
analysis of a document to redefine the importance of terms which is then incorporated into BM25 and statistical language models. 
Kanhabua and N\o rvag \cite{kanhabua2012learning} proposed a time-sensitive ranking model based on learning-to-rank techniques for explicit temporal queries. To learn the ranking model, temporal and entity-based features are applied. 

Regarding more recent works,  Singh et al. \cite{singh2016history} introduced the notion of {\em Historical Query Intents} and modeled it as a search result diversification task which intends to present the most relevant results from a topic-temporal space. For retrieving and ranking historical documents like news articles, the authors propose a retrieval algorithm, called HistDiv, which jointly considers the dimensions of aspect and time. 
Expedition \cite{singh2016expedition} is a time-aware search system for scholars which allows users to search articles in a news collection by entering free-text queries and choosing from four retrieval models: Temporal Relevance, Temporal Diversity, Topical Diversity, and Historical Diversity.
Tempas \cite{holzmann2016tempas} is a search system for web archives that exploits a social bookmarking service (Delicious) for temporally searching an archive by indexing tags and time. The new version of Tempas \cite{holzmann2017exploring} exploits temporal link graphs and the corresponding anchor texts. The authors show how temporal anchor texts can be effective in answering queries beyond purely navigational intents, like finding the most central web pages of an entity in a given time period. 

{\em Difference of our case.}
Similar to the above works, our objective is to rank documents for a user information need. However, our case has the following three distinctive characteristics: 
\begin{itemize}
    \item[i)] The full contents of the documents are not available and there are no term-based indexes on top of them. We have access only on the RDF triples existing in the semantic layer, i.e., on metadata about the documents and on the entities mentioned in the documents. Moreover, these entities have been extracted using automated entity linking systems and thus are prone to disambiguation errors. 
    \item[ii)] The information needs are expressed through structured SPARQL queries, not keywords. These SPARQL queries request documents of a specific time period mentioning one or more specific entities, while these query entities are specified through URIs which means that there is no ambiguity about them. 
    \item[iii)] We already know the documents that match the query, however there are no relevance scores, i.e., all documents {\em equally} match the query. Our objective is to identify (and rank higher) the documents that discuss important information about the entities given in the query. 
\end{itemize}

\subsubsection{Ranking in Knowledge Graphs}
\label{subsec:graph_ranking}
   
There is a plethora of works on ranking entities, concepts and resources in knowledge graphs \cite{Ngonga17holistic,campinas2012effective,hogan2011searching}, as well as on ranking the results returned by SPARQL queries \cite{feyznia2014colina,latifi2014query,mulay2011spring}. The majority of these works exploits the structure of the graph and applies some variation of a popular link analysis algorithm. The survey in \cite{roa2014survey} formalizes and contextualizes the problem of ranking in the Web of Data and provides an analysis and contrast of the similarities, differences and applicability of the different approaches. 

An interesting related line of research tackles the problem of {\em ad-hoc object retrieval} \cite{pound2010ad, tonon2012combining}. In this problem, the input is a keyword query and the output is one or more resources (entity URIs) that satisfy the corresponding information need.
To tackle this kind of problem, Pound et al. \cite{pound2010ad} propose an adaptation of TF-IDF, while Tonon et al. \cite{tonon2012combining} combine an inverted index with entity graph traversal. 
In the same context, the SemSearch challenge \cite{halpin2010evaluating} focused on finding the entity identifier of a specific entity described by a user query, while the TREC entity track \cite{balog2010overview} studied two related search tasks: i) finding all entities related to a given entity, and ii) finding entities with common properties given some examples.

{\em Difference of our case.}
In all these works, the result is a ranked list of {\em well-structured} RDF resources (like entities, properties, or triples). On the contrary, in our problem the units of retrieval represent unstructured (textual) documents returned by structured (SPARQL) queries.
These works operate over knowledge graphs like DBpedia, where entities are described through properties and associations with other entities. A semantic layer is a special kind of a knowledge graph that represents metadata and annotation information about a collection of textual documents like news articles. Given a non-ambiguous SPARQL query and its result (the documents that match the query), our aim is to identify those documents that discuss important information about the query entities in the requested time-period by exploiting only the contents of the semantic layer. 

\section{Problem Definition and Evaluation Dataset}
\label{sec:problemDef}

In this section, we formalize the problem of ranking documents returned by structured queries and describe a new ground truth dataset for the problem at hand. 
First we introduce the required notions and notations.

\subsection{Notions and Notations}

\subsubsection*{Entities.}
In our problem, an {\em entity} is anything with a separate and meaningful existence that also has an identity expressed through a reference in a knowledge base (e.g., a Wikipedia/DBpedia URI). This does not only include persons, locations, organizations, etc., but also events (e.g., {\em US 2016 presidential election}) and more abstract concepts such as {\em democracy} or {\em abortion}. Let $E$ be a finite set of entities, e.g., all Wikipedia entities, where each entity $e \in E$ is associated with a unique URI in the reference knowledge base.

{\em Documents and extracted entities.}
Let $D$ be a set of documents (e.g., a set of news articles) published within a set of time periods $T_D$ of fixed granularity $\Delta$ (e.g., day). For a document $d \in D$, let $t_d \in T_D$ be the time period of granularity $\Delta$ in which $d$ was published, while for a time period $t \in T_D$, let $docs(t) \subseteq D$ be the set of all documents published within $t$, i.e., $docs(t) = \{d \in D ~|~ t_d = t\}$. Let also  $ents(d) \subseteq E$ be all entities mentioned in $d$ extracted using an entity linking system \cite{shen2015entity}.
Inversely, for an entity $e \in E$, let $docs(e) \subseteq D$ be all documents that mention $e$, i.e., $docs(e) = \{d \in D ~|~ e \in ents(d)\}$.

\subsection{Problem Definition}
Given a corpus of documents $D$,
a set of entities $E_D \in E$ mentioned in documents of $D$,
and a SPARQL query $Q$ requesting documents from $D$
published within a {\em time period} $T_Q \subseteq T_D$ and
related to one or more query entities $E_Q \subseteq E_D$
with logical {\tt AND} (mentioning all the query entities) or {\tt OR} (mentioning at least one of the query entities) semantics,
the problem is how to rank the documents $D_Q \subseteq D$ that match $Q$.

Listing \ref{fig:modelingExampleQ1} shows an example SPARQL query
requesting documents published in 1990
discussing about the entities {\em Nelson Mandela}
and {\em Frederik Willem de Klerk} 
(logical {\tt AND} semantics),
while the query in Listing \ref{fig:modelingExampleQ2} requests
articles of 1990 discussing about {\em state presidents of South Africa}
(logical {\tt OR} semantics).
Our objective is to rank the results returned by such SPARQL queries.

\begin{figure}[th]
\centering \scriptsize
\begin{Verbatim}[frame=lines,numbers=left,numbersep=1pt]
 SELECT DISTINCT ?article WHERE {
   ?article dc:date ?date FILTER(year(?date) = 1990) .
   ?article oae:mentions ?entity1, ?entity2 .
   ?entity1 oae:hasMatchedURI  dbr:Nelson_Mandela .
   ?entity2 oae:hasMatchedURI  dbr:F._W._de_Klerk }
\end{Verbatim}
\vspace{-4mm}
\caption{SPARQL query for retrieving articles of 1990 discussing
about {\em Nelson Mandela} and {\em Frederik Willem de Klerk} (logical {\tt AND} semantics).}
\label{fig:modelingExampleQ1}

\centering \scriptsize
\begin{Verbatim}[frame=lines,numbers=left,numbersep=1pt]
 SELECT DISTINCT ?article WHERE {
   SERVICE <http://dbpedia.org/sparql> {
     ?p dc:subject dbc:State_Presidents_of_South_Africa> }
   ?article dc:date ?date FILTER(year(?date) = 1990) .
   ?article oae:mentions ?entity .
   ?entity oae:hasMatchedURI  ?p }
\end{Verbatim}
\vspace{-4mm}
\caption{SPARQL query for retrieving articles of 1990 discussing
about {\em state presidents of South Africa} (logical {\tt OR} semantics).}
\label{fig:modelingExampleQ2}
\vspace{-3mm}
\end{figure}

\subsection{Evaluation dataset}
\label{subsec:evalDataset}

Due to the lack of benchmark datasets for our problem, and to enable empirical evaluations, we have created a new ground truth dataset.
We used the New York Times (NYT) annotated corpus \cite{sandhaus2008new} as the underlying document collection. The corpus contains over 1.8 million articles published by NYT between 1987 and 2007. We used Babelfy \cite{moro2014entity} for extracting DBpedia entities from each article, using a configuration proposed by the Babelfy developers\footnote{The configuration is available at: \url{https://github.com/dice-group/gerbil/blob/master/src/main/java/org/aksw/gerbil/annotator/impl/babelfy/BabelfyAnnotator.java}}. We tested this configuration in the AIDA/CONLL-Test B ground truth dataset \cite{hoffart2011robust} and got the following evaluation scores: micro precision: 0.818, micro recall: 0.684, micro F1: 0.745 (the accuracy is almost the same with the one reported in the Babelfy paper for the same dataset \cite{moro2014entity}). 
Based on these annotations, we constructed a semantic layer following the process described in \cite{fafalios2017SemLayer}. 
Then, we created 24 SPARQL queries, each one requesting articles published in a specific {\em time period} and mentioning one or more {\em entities}. The queries are grouped into 4 categories:  

\begin{itemize}
\item   \textbf{Single-entity queries (Q1-Q6):} 6 queries requesting articles related to 1 entity (e.g., articles of 1990 discussing about {\em Nelson Mandela}).

\item   \textbf{Multiple-entity {\bf\tt AND} queries (Q7-Q12):} 6 queries requesting articles related to 2 or more entities with logical {\tt AND} semantics (e.g., articles of 1990 discussing about {\em Nelson Mandela} \underline{and} {\em F.W. de Klerk}).

\item   \textbf{Multiple-entity {\bf\tt OR} queries (Q13-Q18):} 6 queries requesting articles related to 2 or more entities with logical {\tt OR} semantics (e.g., articles of 1990 discussing about {\em Nelson Mandela} \underline{or} {\em F.W. de Klerk}).

\item   \textbf{Category queries (Q19-Q24):} 6 queries requesting articles related to entities belonging to a DBpedia category (e.g., articles of 1990 discussing about {\em presidents of South Africa\footnote{Entities that have the value \url{<http://dbpedia.org/resource/Category:Presidents_of_South_Africa>} in their subject property (\url{<http://purl.org/dc/terms/subject>}).}}). This category is a special case of multiple-entity {\tt OR} queries where the number of entities can be very large (hundreds or even thousands).
\end{itemize}
We manually evaluated all the results returned by these queries (773 results totally) using a graded relevance scale (from 0 to 3), following the criteria described below:

\begin{itemize}
\item   \textbf{Score 0}: The document has almost nothing to do with the query entities.
\item   \textbf{Score 1}: The topic of the document is \textbf{not} about the query entities, however the query entities are related to the document context.
\item   \textbf{Score 2}: The topic of the document is \textbf{not} about the query entities, however the query entities are important for the document context.
\item   \textbf{Score 3}: The topic of the document is about the query entities and discusses something important about them.
\end{itemize}

Table \ref{tab:queriesAndRelevance} shows the number of results per relevance score for each of the queries. 
The semantic layer, the SPARQL queries, and the relevance scores (together with explanations for the provided scores) are publicly available\footnote{\url{http://l3s.de/~fafalios/jcdl/evaluation_dataset.zip}}.

\begin{table}
  \caption{Number of results per relevance score for each query of the evaluation dataset.}
   \vspace{-2mm}
   \renewcommand{\arraystretch}{0.75}
  \label{tab:queriesAndRelevance}
  \begin{tabular}{cccccc}
    \toprule
    \makecell{Query} &
    \makecell{Total number\\of results} &
    \makecell{Score 0\\results} &
    \makecell{Score 1\\results} &
    \makecell{Score 2\\results} & 
    \makecell{Score 3\\results}\\
    \midrule
     1 & 65 & 13 & 13 & 4 & 35 \\
     2 & 28 & 20 & 5 & 1 & 2 \\
     3 & 28 & 24 & 2 & 0 & 2 \\
     4 & 23 & 15 & 3 & 1 & 4 \\
     5 & 38 & 18 & 6 & 7 & 7 \\
     6 & 24 & 16 & 4 & 1 & 3 \\
     \midrule
     7 & 29 & 13 & 6 & 8 & 2 \\
     8 & 61 & 10 & 27 & 15 & 9 \\
     9 & 42 & 14 & 10 & 10 & 8 \\
     10 & 28 & 17 & 4 & 6 & 1 \\
     11 & 27 & 13 & 10 & 4 & 0 \\
     12 & 24 & 13 & 2 & 8 & 1 \\
     \midrule
     13 & 30 & 21 & 1 & 4 & 4 \\
     14 & 37 & 15 & 8 & 8 & 6 \\
     15 & 27 & 18 & 5 & 2 & 2 \\
     16 & 23 & 5 & 14 & 2 & 2 \\
     17 & 21 & 13 & 2 & 2 & 4 \\
     18 & 25 & 6 & 4 & 9 & 6 \\
     \midrule
     19 & 41 & 24 & 4 & 8 & 5 \\
     20 & 22 & 11 & 8 & 3 & 0 \\
     21 & 29 & 16 & 5 & 5 & 3 \\
     22 & 47 & 33 & 7 & 3 & 4 \\
     23 & 31 & 18 & 3 & 8 & 2 \\
     24 & 23 & 14 & 3 & 3 & 3 \\
     \bottomrule
\end{tabular}
\vspace{-2mm}
\end{table}

\section{Probabilistic Modeling}
\label{sec:probabilisticModeling}

The question is: {\em "what makes an archived document important given a time period and one or more query entities"?}
We focus on an approach that makes use of the least amount of metadata information about the archived documents. This includes the {\em publication date} and the {\em extracted entities} of each document.
Given this data, we have identified the following aspects that can affect the importance of a document to a query: 

\begin{itemize}
\item the {\em relativeness} of a document with respect to the query entities (the document should talk about the query entities, ideally as its main topic).
\item the {\em timeliness} of a document with respect to its publication date (the document should have been published in a time period which is important for the query entities).
\item the {\em relatedness} of a document with respect to its reference to other entities (the document should discuss the relation of the query entities with other entities that are important for the query entities in important time periods).
\end{itemize}

The idea is to promote documents that:
i) mention the query entities many times in their contents (because then, the topic of the document may be about these entities),
ii) have been published in important (for the query entities) time periods, and 
iii) mention many other entities that co-occur frequently with the query entities in important time periods.
For example, in case we want to rank articles of 1990 discussing about {\em Nelson Mandela}, we want to favor articles that
i) discuss about {\em Nelson Mandela} as their main topic,
ii) have been published in important (for {\em Nelson Mandela}) time periods (e.g., February of 1990 since during that period he was released from prison), and
iii) mention other entities that seem to be important for Nelson Mandela during important time periods (e.g., {\em Frederik Willem de Klerk} who was South Africa's State President in 1990).

\subsection{Relativeness}

We consider that if the query entities are mentioned multiple times
within a document, the document should receive a high score since the document's topic may be about these entities. The term frequency (in our case entity frequency) is a classic numerical statistic that is intended to reflect how important a word (entity) is to a document \cite{leskovec2014mining}.

We first define a {\em relativeness} score of a document $d \in D_Q$ based on the {\em frequency} of the query entities in $d$. First, let $count(e, d)$ be the number of occurrences of $e$ in document $d$. For the case of {\tt AND} semantics (\q{$\wedge$}), the score is defined as:

\begin{equation}
score^{f}_{\wedge}(d, E_Q) = \frac{\sum_{e \in E_Q}{count(e, d)}}{\sum_{e' \in ents(d)}{count(e', d)}}
\end{equation}

Notice that the score of a document will be $1.0$ if it contains the query entities and no other entity.
For the case of {\tt OR} semantics (\q{$\vee$}), we can also consider the number of query entities mentioned in the document (since a document does not probably contain all the query entities as in the case of {\tt AND} semantics). In this case, the {\em relativeness} score can be defined as follows:

\begin{equation}
score^{f}_{\vee}(d, E_Q) = \frac{\sum_{e \in E_Q}{count(e, d)}}{\sum_{e' \in ents(d)}{count(e', d)}} \cdot \frac{|ents(d) \cap E_Q|}{|E_Q|}
\end{equation}
where $\frac{|ents(d) \cap E_Q|}{|E_Q|}$ is the percentage of query entities discussed in the document.
The score of a document will be 1.0 if it contains all the query entities and no other entity.
This formula favors documents mentioning many of the query entities multiple times.

Now, the probability of a retrieved document $d \in D_Q$ given only the query entities can be defined as:
\begin{equation}
P(d|E_Q) = \frac{score^{f}(d, E_Q)}{\sum_{d' \in D_Q}{score^{f}(d', E_Q)}}
\end{equation}

\subsection{Timeliness}
Previous works on searching document archives have shown that, considering the fraction of documents published in a time period and mentioning the query entities can improve the effectiveness of document retrieval (compared to an approach that assumes equal distribution) \cite{singh2016history}. 
In a similar way, and for the case of {\tt AND} semantics, we define the following importance score of a {\em time period} $t \in T_Q$ :
\begin{equation}
score^{t}_\wedge(t) = \frac{|docs(t) \cap D_Q|}{|D_Q|}
\end{equation}

This scoring formula favors time periods in which there is a large number of documents discussing about the query entities.

For the case of {\tt OR} semantics, in a time period $t$ there may be a large number of documents discussing only for one of the query entities, while in another time period $t'$ there may be a smaller number of documents discussing though for many of the query entities.
For also taking into account the number of query entities discussed in documents of a specific time period, we consider the following formula:
\begin{equation}
score^{t}_\vee(t) = \frac{|docs(t) \cap D_Q|}{|D_Q|} \cdot N(E_Q,t)
\end{equation}
where, $N(E_Q,t)$ is the average percentage of query entities discussed in articles of $t$, i.e.:
\begin{equation}
N(E_Q,t) =  \frac{\sum_{d \in docs(t) \cap D_Q}{\frac{|ents(d) \cap E_Q|}{|E_Q|}}}{|docs(t) \cap D_Q|}
\end{equation}

Now, the probability of a retrieved document $d \in D_Q$ given only its publication date $t_d$ can be defined as:

\begin{equation}
P(d | t_d) = \frac{score^{t}(t_d)}{\sum_{d' \in D_Q}{score^{t}(t_{d'})}}
\end{equation}

\subsection{Relatedness}

Recent works have shown that the co-occurrence of entities in documents of a specific time period is a strong indicator of their relatedness during that period \cite{zhang2016probabilistic,tran2017beyond}. 
We also consider that entities that are co-mentioned frequently with the query entities in important time periods are probably important for them.
However, there may be also some general entities that co-occur with the query entities in almost all documents (independently of the time period).
Thus, we should also avoid over-emphasizing documents mentioning such \q{common} entities.

For the case of {\tt AND} semantics, we consider the following {\em relatedness} score
for an entity $e \in E_D \setminus E_Q$:

\begin{equation}
\begin{split}
score^{r}_\wedge(e) = & ~ idf_\wedge(e) \cdot \sum_{t \in T_Q}{(score^{t}_\wedge(t) \cdot \frac{|docs(t) \cap D_Q \cap docs(e)|}{|docs(t) \cap D_Q|})}\\
= & ~ idf_\wedge(e) \cdot \sum_{t \in T_Q}{\frac{|docs(t) \cap D_Q \cap docs(e)|}{|D_Q|}}
\end{split}
\end{equation}
where $idf_\wedge(e)$ is the inverse document frequency
of entity $e$ in the set of documents discussing about the query entities in the entire corpus,
which can be defined as follows:
\begin{equation}
\label{frml:idf_and}
idf_\wedge(e) = 1 - \frac{|docs(e) \cap (\cap_{e' \in E_Q}{docs(e')})|}{|\cap_{e' \in E_Q}{docs(e')}|}
\end{equation}

The formula considers the percentage of
documents in which the entity
co-occurs with the query entities in important time periods.

For the case of {\tt OR} semantics,
the above formula does not consider the
number of different query entities discussed in documents together with the entity $e$.
To also handle this aspect,
we consider the following {\em relatedness} score for the case of {\tt OR} semantics:

\begin{equation}
\begin{split}
score^{r}_\vee(e) =  & ~ idf_\vee(e)~N(E_Q, e) \sum_{t \in T_Q}{(score^{t}_\vee(t) \frac{|docs(t) \cap D_Q \cap docs(e)|}{|docs(t) \cap D_Q|})} \\
= & ~ idf_\vee(e)~N(E_Q, e) \sum_{t \in T_Q}{(N(E_Q, t) \frac{|docs(t) \cap D_Q \cap docs(e)|}{|D_Q|})}
\end{split}
\label{eq:relatednessScore}
\end{equation}
where $N(E_Q, e)$ is
the average percentage of query entities discussed in articles together with entity $e$, i.e.:
\begin{equation}
N(E_Q, e) =  \frac{\sum_{d \in docs(e) \cap D_Q}{\frac{|ents(d) \cap E_Q|}{|E_Q|}}}{|docs(e) \cap D_Q|}
\end{equation}

Now the inverse document frequency $idf_\vee(e)$
includes documents mentioning at least one of the query entities, i.e.:
\begin{equation}
\label{frml:idf_or}
idf_\vee(e) = 1 - \frac{|docs(e) \cap (\cup_{e' \in E_Q}{docs(e')})|}{|\cup_{e' \in E_Q}{docs(e')}|}
\end{equation}

This formula favors related entities that
i)  co-occur frequently with many of the query entities,
ii) are discussed in documents published in important (for the query entities) time periods.

Now, the probability of a document $d \in D_Q$
given only other entities mentioned in the retrieved documents ($E_{D_Q}$) can be defined as:
\begin{equation}
P(d | E_{D_Q}) = \frac{\sum_{e \in ents(d) \setminus E_Q}{score^{r}(e)}}
                  {\sum_{d' \in D_Q}{\sum_{e' \in ents(d') \setminus E_Q}{score^{r}(e')}}}
\end{equation}

\subsection{Joining the Models}

We can now combine the different models in a single probability score:
\begin{equation}
P(d | E_Q, t_d, E_{D_Q}) = \frac{P(d | E_Q)P(d | T_Q)P(d | E_{D_Q})}{\sum_{d' \in D_Q}{P(d' | E_Q)P(d' | T_Q)P(d' | E_{D_Q})}}
\end{equation}
where the denominator can be ignored as it does not influence the ranking.

\section{Stochastic Modeling}
\label{sec:stochasticModeling}

Here we model the problem as a {\em random walker} on the graph ({\em Markov chain}) defined by the query-entities $E_Q$, the returned documents $D_Q$, and the entities mentioned in the documents $E_{D_Q}$. Then, we propose a biased (personalized-like) PageRank algorithm for analyzing the graph and scoring its nodes.
Similar modeling and scoring methods have been applied for the problems of {\em results re-ranking} \cite{fafalios2017jasist} and {\em enrichment} \cite{fafalios2014postAnalysis} in information retrieval.

\subsection{The Transition Graph}
The walker starts from a query-entity and can move either to a document mentioning the entity or to a related entity that co-occurs with the query-entity in at least one document. From a document, the walker can move to an entity mentioned in it, while from a no query-entity, the walker can only move to a document mentioning that entity.  

In the case of logical {\tt AND} semantics, the query-entities are connected with all documents, while in the case of {\tt OR} semantics, each query-entity is connected with at least one document. 
Figure \ref{fig:graphexample} shows an example of a transition graph for the case of {\tt OR} semantics. In this example, the query-entities are two (the black nodes), while we notice that three of the documents mention both query-entities ($d_2$, $d_3$ and $d_4$). 

\renewcommand{\figurename}{Figure}
\setcounter{figure}{1}
\begin{figure}
\centering
\fbox{\includegraphics[width=1.6in]{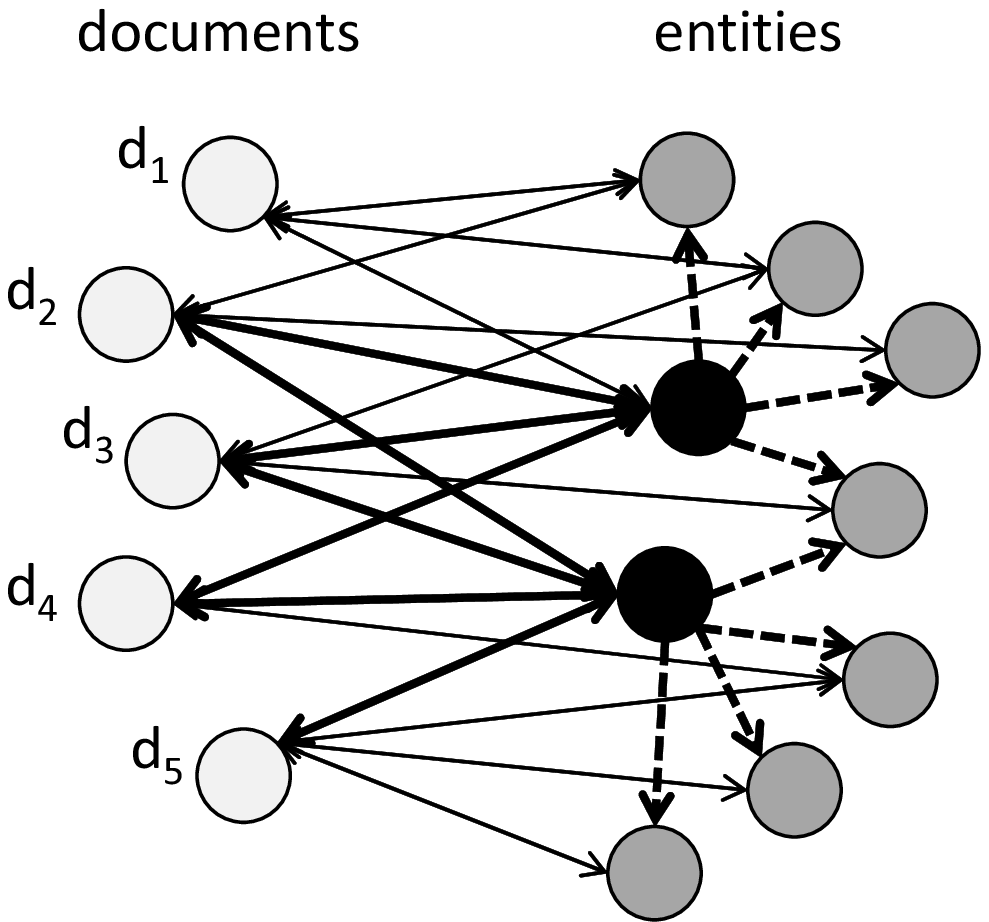}}
\vspace{-2mm}
\caption{An example of the considered transition graph for the case of logical {\tt OR} semantics (with the query-entities being the black nodes).}
\label{fig:graphexample}
\vspace{-4mm}
\end{figure}

\subsection{The Transition Probabilities}

\subsubsection*{From query-entities to documents or related entities}

When the walker lies at a query-entity $e \in E_Q$, he can either move to a document $d$ mentioning the entity or to a related entity $e'$. When moving to a document, we consider its {\em relativeness} and {\em timeliness} scores as introduced in the previous section. Specifically, the weight of the edge from a query-entity $e$ to a document $d$ is defined as:  

\begin{equation}
\label{eq:trans1}
\begin{split}
weight(e \rightarrow d) = \frac{score^{f}(d,E_Q) \cdot score^{t}(t_d)}{\sum_{d' \in docs(e) \cap D_Q}(score^{f}(d',E_Q) \cdot score^{t}(t_{d'}))}
\end{split}
\end{equation}

For moving from a query-entity $e$ to a related entity $e'$, we consider the {\em relatedness} score of $e'$:

\begin{equation}
\label{eq:trans2}
\begin{split}
weight(e \rightarrow e') = \frac{score^{r}(e')}{\sum_{e'' \in out(e)}(score^{r}(e''))}
\end{split}
\end{equation}
where $out(e)$ is the set of nodes connected to $e$ through outgoing edges starting from $e$. 

However, the weights of the outgoing edges of a single node represent transition
probabilities and must sum to 1.
Thus, the weight of the edge from a query-entity $e$ to a connected node $n$ (that can be either a document or a related entity) is defined as: 

\begin{gather}
\label{eq:queryentity2node}
weight(e \rightarrow n) = \begin{cases}
    p_1 \cdot weight(e \rightarrow d) \qquad
    n \in D_Q\\
    (1-p_1) \cdot weight(e \rightarrow e') \qquad n \in E_{D_Q}
\end{cases}
\end{gather}
where $p_1 \in [0,1]$ is the probability that the walker selects to move to a document node.  

\subsubsection*{From documents to entities}

For moving from a document $d$ to an entity $e$ mentioned in $d$, it is more likely that the walker moves to an entity that is mentioned many times within its contents. Thus, we simply consider the normalized frequency of $e$ in $d$. Specifically:  
\begin{equation}
\label{eq:trans4}
weight(d \rightarrow e) = \frac{count(e, d)}{\sum_{e' \in ents(d)}(count(e',d))}
\end{equation}

\subsubsection*{From no query-entities to documents}

Likewise, when moving from a no query-entity $e$ to a document $d$, it is more likely that the walker moves to a document that mentions $e$ many times. Thus, we again consider the normalized frequency of $e$ in $d$. Specifically:  

\begin{equation}
\label{eq:trans5}
weight(e \rightarrow d) = \frac{count(e,d)}{\sum_{d' \in docs(e) \cap D_Q}(count(e,d')}
\end{equation}

\subsection{The Stochastic Analysis (Random Walk with Restart)}
For analyzing the transition graph, we follow a PageRank-like algorithm. The walker starts from the query-entities and can either follow an edge or perform a \q{restart}, i.e., jump to a query-entity and start again the traversal . Formally, the score of a graph node $n$ is defined as:

\begin{equation}
\label{eq:prAlg}
r(n) = d \cdot Jump(n) + (1-d) \cdot \sum_{n' \in in(n)}{(weight(n' \rightarrow n) \cdot r(n'))}
\end{equation}
where $d\in [0,1]$ is the probability that the walker performs a restart, $Jump(n)$ is the probability the walker to restart by jumping to node $n$, $in(n)$ is the set of nodes connected to $n$ through incoming edges, and $weight(n \rightarrow n')$ (as defined in Formulas \ref{eq:trans1}-\ref{eq:trans5}) is the probability that the walker visits $n$ when being at node $n'$ connected to $n$ (there should be an edge from $n'$ to $n$). 

As regards the {\em restart}, we allow equiprobable jumps only to query entities, i.e.:

\begin{equation}
\label{eq:jump}
Jump(n) = \left\{
\begin{array}{l l}
    1 / |E_Q| & \quad n \in E_Q \\
    0 & \quad n \notin E_Q\\
\end{array} \right.
\end{equation}

Regarding the initial scores of the graph nodes, we assign the same score to the query entities ($1 / |E_Q|$), while the score of all other entities is zero. Finally, the algorithm must iteratively run to convergence. 

\section{Evaluation}
\label{sec:evaluation}

We evaluated the performance of the proposed models using the ground truth dataset described in Section \ref{subsec:evalDataset} and two evaluation measures: {\em Normalized Discounted Cumulative Gain (NDCG)} at positions 5, 10, full list, and {\em Precision (P)} at 5 and 10 (precision for full list is the same in all cases). Precision is the fraction of relevant results among the top retrieved results, while NDCG uses a graded relevance scale and measures the usefulness, or gain, of a document based on its position in the result list \cite{jarvelin2002cumulated}.
For precision, we consider a document as relevant if its relevance score is either 2 or 3 and irrelevant if it is either 0 or 1. As regards timeliness, we considered {\em day} granularity. 
We also tested the case of random rankings of the results. In this case, we computed 10 different random lists for each query and considered the average NDCG and precision scores. 

\subsection{Effectiveness of Probabilistic Models}
\label{subsec:evalProbModels}
We investigated the performance of each model described in Section \ref{sec:probabilisticModeling} as well as their combinations. 

\subsubsection{Overall Results}

Table \ref{tab:ndcg_all} shows the average NDCG and precision scores for all queries.
We notice that joining all three models provides the best results. The improvement of the top-5 results compared to {\em relativeness} is about 17\% in NDCG and 18\% in precision, which is statistically significant (paired t-test, $p \leq 0.05$). Note that {\em relativeness} can be considered a baseline model for our problem since it considers entity frequency which is a classic numerical statistic \cite{leskovec2014mining}. The results also show that {\em relatedness} performs very well, outperforming the joined models for P@10. This means that considering other entities that co-occur frequently with the query entities in important time periods has a positive effect on the ranking.

\begin{table*}
\setlength\tabcolsep{3.0pt}
  \caption{Average NDCG and Precision of the probabilistic models for all queries.}
   \vspace{-3.5mm}
  \label{tab:ndcg_all}
  \renewcommand{\arraystretch}{0.75}
  \begin{tabular}{l|cccccccc}
    \toprule
    \makecell{Measure} & 
    \makecell{Random\\ranking} &
    \makecell{Relativeness\\{[}A{]}} &
    \makecell{Timeliness\\{[}B{]}} &
    \makecell{Relatedness\\{[}C{]}} &
    \makecell{{[}A{]}{[}B{]}} &
    \makecell{{[}A{]}{[}C{]}} &
    \makecell{{[}B{]}{[}C{]}} &
    \makecell{{[}A{]}{[}B{]}{[}C{]}}\\
    \midrule
    NDCG@5  & 0.26 & 0.48 & 0.27 & 0.41 & 0.53 & 0.52 & 0.45 & {\bf 0.56} \\
    NDCG@10 & 0.33 & 0.52 & 0.36 & 0.50 & 0.55 & 0.56 & 0.51 & {\bf 0.58} \\
    NDCG@all & 0.68 & 0.79 & 0.69 & 0.76 & 0.80 & 0.81 & 0.76 & {\bf 0.82} \\
    \midrule
    P@5 & 0.27 & 0.44 & 0.28 & 0.48 & 0.48 & 0.50 & 0.45 & {\bf 0.52}  \\
    P@10 & 0.29 & 0.38 & 0.30 & {\bf 0.45} & 0.37 & 0.42 & 0.41 & 0.41 \\
    \bottomrule
\end{tabular}
\end{table*}

\begin{table}
\setlength\tabcolsep{2.8pt}
  \caption{Average NDCG and Precision of the probabilistic models for single-entity queries (Q1-Q6).}
  \vspace{-3.5mm}
  \label{tab:ndcg_1}
  \renewcommand{\arraystretch}{0.75}
  \begin{tabular}{c|cccccccc}
    \toprule
    \makecell{Measure} & 
    \makecell{{[}A{]}} &
    \makecell{{[}B{]}} &
    \makecell{{[}C{]}} &
    \makecell{{[}A{]}{[}B{]}} &
    \makecell{{[}A{]}{[}C{]}} &
    \makecell{{[}B{]}{[}C{]}} &
    \makecell{{[}A{]}{[}B{]}{[}C{]}}\\
    \midrule
    NDCG@5 & 0.66 & 0.30 & 0.40 & 0.68 & 0.69 & 0.45 & {\bf 0.70} \\
    NDCG@10 & 0.69 & 0.38 & 0.51 & 0.70 & {\bf 0.71} & 0.51 & 0.70 \\
    NDCG@all & {\bf 0.88} & 0.67 & 0.75 & 0.87 & {\bf 0.88} & 0.72 & 0.87 \\
    \midrule
    P@5 & 0.57 & 0.23 & 0.50 & 0.57 & {\bf 0.60} & 0.40 & {\bf 0.60}  \\
    P@10 & 0.40 & 0.27 & {\bf 0.45} & 0.40 & 0.40 & 0.38 & 0.40 \\
    \bottomrule
\end{tabular}
\end{table}

\subsubsection{Detailed results per query type.}
Tables \ref{tab:ndcg_1}-\ref{tab:ndcg_4} show the results per query type.

\begin{table}
\setlength\tabcolsep{2.8pt}
  \caption{Average NDCG and Precision of the probabilistic models for multiple-entity AND queries (Q7-Q12).}
  \vspace{-3.5mm}
  \label{tab:ndcg_2}
  \renewcommand{\arraystretch}{0.75}
  \begin{tabular}{c|cccccccc}
    \toprule
    \makecell{Measure} & 
    \makecell{{[}A{]}} &
    \makecell{{[}B{]}} &
    \makecell{{[}C{]}} &
    \makecell{{[}A{]}{[}B{]}} &
    \makecell{{[}A{]}{[}C{]}} &
    \makecell{{[}B{]}{[}C{]}} &
    \makecell{{[}A{]}{[}B{]}{[}C{]}}\\
    \midrule
    NDCG@5  & 0.34 & 0.28 & 0.31 & {\bf 0.42} & 0.35 & 0.38 & 0.40 \\
    NDCG@10  & 0.43 & 0.33 & 0.40 & 0.46 & 0.45 & 0.46 & {\bf 0.47} \\
    NDCG@all  & 0.76 & 0.71 & 0.75 & {\bf 0.78} & 0.76 & 0.77 & 0.77 \\
    \midrule
    P@5 & 0.43 & 0.33 & 0.30 & {\bf 0.50} & 0.47 & 0.43 & {\bf 0.50}  \\
    P@10 & 0.50 & 0.32 & 0.42 & 0.48 & {\bf 0.53} & 0.47 & 0.50 \\
    \bottomrule
\end{tabular}
\end{table}

\begin{table}
\setlength\tabcolsep{2.8pt}
  \caption{Average NDCG and Precision of the probabilistic models for multiple-entity OR queries (Q13-Q18).}
  \vspace{-3.5mm}
  \label{tab:ndcg_3}
  \renewcommand{\arraystretch}{0.75}
  \begin{tabular}{c|cccccccc}
    \toprule
    \makecell{Measure} & 
    \makecell{{[}A{]}} &
    \makecell{{[}B{]}} &
    \makecell{{[}C{]}} &
    \makecell{{[}A{]}{[}B{]}} &
    \makecell{{[}A{]}{[}C{]}} &
    \makecell{{[}B{]}{[}C{]}} &
    \makecell{{[}A{]}{[}B{]}{[}C{]}}\\
    \midrule
    NDCG@5 & 0.68 & 0.24 & 0.44 & 0.70 & {\bf 0.72} & 0.46 & 0.71 \\
    NDCG@10  & 0.69 & 0.36 & 0.55 & 0.69 & 0.72 & 0.54 & {\bf 0.73} \\
    NDCG@all  & 0.87 & 0.69 & 0.79 & 0.87 & {\bf 0.88} & 0.79 & {\bf 0.88}  \\
    \midrule
    P@5 & 0.60 & 0.27 & 0.57 & 0.60 & {\bf 0.63} & 0.43 & {\bf 0.63}  \\
    P@10 & 0.42 & 0.30 & {\bf 0.48} & 0.40 & 0.45 & 0.42 & 0.45 \\
    \bottomrule
\end{tabular}
\end{table}

\begin{table}
\setlength\tabcolsep{2.8pt}
  \caption{Average NDCG and Precision of the probabilistic models for category queries (Q19-Q22).}
  \vspace{-3.5mm}
  \label{tab:ndcg_4}
  \renewcommand{\arraystretch}{0.75}
  \begin{tabular}{c|cccccccc}
    \toprule
    \makecell{Measure} & 
    \makecell{{[}A{]}} &
    \makecell{{[}B{]}} &
    \makecell{{[}C{]}} &
    \makecell{{[}A{]}{[}B{]}} &
    \makecell{{[}A{]}{[}C{]}} &
    \makecell{{[}B{]}{[}C{]}} &
    \makecell{{[}A{]}{[}B{]}{[}C{]}}\\
    \midrule
    NDCG@5 & 0.22 & 0.27 & 0.48 & 0.34 & 0.34 &  {\bf 0.52} & 0.41 \\
    NDCG@10 & 0.28 & 0.37 & {\bf 0.53} & 0.36 & 0.38 &  {\bf 0.53} & 0.43 \\
    NDCG@all & 0.66 & 0.68 & {\bf 0.78} & 0.70 & 0.71 &  0.77 & 0.73 \\
    \midrule
    P@5 & 0.17 & 0.27 & {\bf 0.53} & 0.23 & 0.30 & {\bf 0.53} & 0.33  \\
    P@10 & 0.20 & 0.30 & {\bf 0.45} & 0.20 & 0.28 & 0.37 & 0.28 \\
    \bottomrule
\end{tabular}
\end{table}

Regarding {\em single entity queries} (Table \ref{tab:ndcg_1}), we observe that combining relativeness with relatedness has almost the same performance with the case of joining all models. This means that timeliness seems to have a minor negative effect on the rankings. 
For this query type, query 4 has the best performance on the joined model with NDCG@5 $=1.0$ and NDCG@10 $=0.97$. On the contrary, query 5 does not perform well with NDCG@5 = 0.29 and NDCG@10 = 0.37 (this query returns several "headlines" articles summarizing what is inside today's paper, which may confuse our modeling).

As regards {\em multiple-entity AND queries} (Table \ref{tab:ndcg_2}), we observe that joining relativeness with timeliness seems to perform better with small difference compared to the case of joining of all models, which means that here timeliness has a positive effect on the ranking. We also notice that all models perform worst compared to the single-entity case. For this query type, the NDCG@10 score of all queries ranges from 0.35 (for query 8) to 0.67 (for query 12).  

The case of {\em multiple-entity OR queries} (Table \ref{tab:ndcg_3}) is very similar to the case of {\em single entity queries} where timeliness seems to have a minor negative impact. Here almost all queries have very good performance with NDCG@10 score $>0.55$.

As regards {\em category queries} (Table \ref{tab:ndcg_4}), it is very interesting that the performance of relativeness is very low, while relatedness performs very well and timeliness has a positive impact especially for the top-5 lists. By thoroughly checking the results of queries that perform very bad in relativeness (like queries 19 and 22), we noticed that the reason for this bad performance is the presence of a large number of disambiguation errors in the returned documents. A characteristic of this type of queries is the big number of query entities which in turn increases the probability of disambiguation errors. Note that relativeness favors documents containing many instances of the query entities. However, if an irrelevant document contains many instances of entity names that have been wrongly linked to query entities, this document will receive a high relativeness score (since it contains many \q{false positive} query entities). For example, query 19 in our dataset has a very bad performance in relativeness (NDCG@5 = 0.02, NDCG@10 = 0.10). This query requests articles mentioning NASA civilian astronauts, however many instances of the entity name {\em Armstrong} have been incorrectly linked to the American astronaut Neil Armstrong (like the referee Hank Armstrong or the company Armstrong World Industries). Likewise, in query 22 several instances of the entity names {\em Teresa} and {\em Theresa} are incorrectly mapped to Mother Teresa.
Nevertheless, we see that {\em relatedness} somehow fixes this problem. This model considers the co-occurrence of entities in important time periods and favors documents that contain many instances of other important entities. The probability that many false positive cases co-occur with the same related entities in many documents is very low, and thus considering entity associations has a positive effect on the ranking. In query 19, for example, the incorrect instances of the query entity Neil Armstrong co-occur with entities that do not exist in the documents in which Neil Armstrong has been correctly identified, and thus the relatedness score (Equation \ref{eq:relatednessScore}) of these entities is low. 

\subsection{Effectiveness of Stochastic Model}

We run experiments for different values of $d$  probability (probability to perform a restart, cf. Equation \ref{eq:prAlg}) and different values of $p_1$ probability (probability to move to a document node when being at a query-entity node, cf. Equation \ref{eq:queryentity2node}).  
For the restart probability, we tested the following 5 cases: $d=0.0, 0.2, 0.4, 0.6, 0.8$. Note that testing $d=1.0$ does not make sense, since the walker will never reach document nodes. Regarding the $p_1$ probability, we tested 6 cases ($p_1=0.0, 0.2, 0.4, 0.6, 0.8, 1.0$). 
Empirically, $d=0.2$ provides the best results independently of the $p_1$ value. Thus, all the results reported below correspond to $d=0.2$. Moreover, in all experiments, we set the number of iterations to 30.  

Tables \ref{tab:ndcg_st_1}-\ref{tab:ndcg_st_4} show the results per query type and for different values of $p_1$ probability.
For the first three types of queries, we observe that $p_1=1.0$ provides the best results and that, as the probability gets lower, the results get worse. Using $p_1=1.0$, the walker can move only to document nodes. This means that for small values of $p_1$ probability, the algorithm overemphasizes the association between the entities resulting in worst performance.  
Thus, the results show that reaching documents through related entities affects negatively the rankings for these types of queries.  

Compared to the results of the probabilistic modeling, we see that: i) the performance of the two models is almost the same for single-entity queries, while the stochastic model outperforms the best probabilistic model for multiple-entity {\tt AND} queries, and ii) the best probabilistic model outperforms the stochastic model for multiple-entity {\tt OR} queries. 
Regarding (ii), this failure of the stochastic model is probably due to the fact that the graph in the case of {\tt OR} queries is not as well-connected as in the case of {\tt AND} queries (for {\tt OR} queries, the graph may contain disconnected components, e.g., in cases where there are no documents mentioning all query entities). 

\begin{table}
\setlength\tabcolsep{2.8pt}
  \caption{Average NDCG and Precision of the stochastic model for single entity queries (Q1-Q6).}
  \vspace{-3.5mm}
  \label{tab:ndcg_st_1}
  \renewcommand{\arraystretch}{0.75}
  \begin{tabular}{c|ccccccc}
    \toprule
    \makecell{Measure} & 
    \makecell{$p_1$=0.0} &
    \makecell{$p_1$=0.2} &
    \makecell{$p_1$=0.4} &
    \makecell{$p_1$=0.6} &
    \makecell{$p_1$=0.8} &
    \makecell{$p_1$=1.0}\\
    \midrule
    NDCG@5   & 0.09 & 0.27 & 0.48 & 0.63 & 0.65 & {\bf 0.67} \\ 
    NDCG@10  & 0.18 & 0.35 & 0.54 & 0.69 & 0.72 & {\bf 0.74} \\ 
    NDCG@all & 0.60 & 0.68 & 0.77 & 0.85 & {\bf 0.87} & {\bf 0.87} \\ 
    \midrule
    P@5 & 0.07 & 0.23 & 0.43 & 0.47 & 0.53 & {\bf 0.60} \\ 
    P@10 & 0.12 & 0.23 & 0.33 & 0.37 & 0.42 & {\bf 0.43} \\ 
    \bottomrule
\end{tabular}
\end{table}

\begin{table}
\setlength\tabcolsep{2.8pt}
  \caption{Average NDCG and Precision of the stochastic model for multiple-entity AND queries (Q7-Q12).}
  \vspace{-3.5mm}
  \label{tab:ndcg_st_2}
  \renewcommand{\arraystretch}{0.75}
  \begin{tabular}{c|ccccccc}
    \toprule
    \makecell{Measure} & 
    \makecell{$p_1$=0.0} &
    \makecell{$p_1$=0.2} &
    \makecell{$p_1$=0.4} &
    \makecell{$p_1$=0.6} &
    \makecell{$p_1$=0.8} &
    \makecell{$p_1$=1.0}\\
    \midrule
    NDCG@5 & 0.29 & 0.29 & 0.32 & 0.31 & 0.37 & {\bf 0.48} \\
    NDCG@10 & 0.36 & 0.36 & 0.40 & 0.45 & 0.47 & {\bf 0.52} \\
    NDCG@all & 0.72 & 0.73 & 0.74 & 0.75 & 0.78 & {\bf 0.80} \\
    \midrule
    P@5 & 0.27 & 0.27 & 0.27 & 0.30 & 0.37 & {\bf 0.57} \\ 
    P@10 & 0.30 & 0.30 & 0.38 & 0.45 & 0.47 & {\bf 0.50} \\ 
    \bottomrule
\end{tabular}
\end{table}

\begin{table}
\setlength\tabcolsep{2.8pt}
  \caption{Average NDCG and Precision of the stochastic model for multiple-entity OR queries (Q13-Q18).}
  \vspace{-3.5mm}
  \label{tab:ndcg_st_3}
  \renewcommand{\arraystretch}{0.75}
  \begin{tabular}{c|ccccccc}
    \toprule
    \makecell{Measure} & 
    \makecell{$p_1$=0.0} &
    \makecell{$p_1$=0.2} &
    \makecell{$p_1$=0.4} &
    \makecell{$p_1$=0.6} &
    \makecell{$p_1$=0.8} &
    \makecell{$p_1$=1.0}\\
    \midrule
    NDCG@5 & 0.20 & 0.30 & 0.32 & 0.38 & 0.45 & {\bf 0.59} \\ 
    NDCG@10 & 0.27 & 0.38 & 0.42 & 0.51 & 0.52 & {\bf 0.65} \\ 
    NDCG@all & 0.66 & 0.71 & 0.72 & 0.75 & 0.76 & {\bf 0.84} \\
    \midrule
    P@5 & 0.27 & 0.33 & 0.33 & 0.37 & 0.47 & {\bf 0.53} \\ 
    P@10 & 0.30 & 0.33 & 0.35 & 0.42 & 0.42 & {\bf 0.45} \\
    \bottomrule
\end{tabular}
\end{table}

\begin{table}
\setlength\tabcolsep{2.8pt}
  \caption{Average NDCG and Precision of the stochastic model for category queries (Q19-Q24).}
  \vspace{-3.5mm}
  \label{tab:ndcg_st_4}
  \renewcommand{\arraystretch}{0.75}
  \begin{tabular}{c|ccccccc}
    \toprule
    \makecell{Measure} & 
    \makecell{$p_1$=0.0} &
    \makecell{$p_1$=0.2} &
    \makecell{$p_1$=0.4} &
    \makecell{$p_1$=0.6} &
    \makecell{$p_1$=0.8} &
    \makecell{$p_1$=1.0}\\
    \midrule
    NDCG@5 & 0.43 & 0.47 & {\bf 0.53} & 0.50 & 0.46 & 0.23 \\ 
    NDCG@10 & 0.48 & 0.54 & {\bf 0.55} & 0.52 & 0.54 & 0.36 \\ 
    NDCG@all & 0.74 & 0.77 & {\bf 0.79} & 0.77 & 0.77 & 0.68 \\ 
    \midrule
    P@5 & 0.57 & 0.57 & {\bf 0.60} & {\bf 0.60} & {\bf 0.60} & 0.23 \\ 
    P@10 & 0.42 & {\bf 0.43} & 0.40 & 0.40 & {\bf 0.43} & 0.28 \\ 
    \bottomrule
\end{tabular}
\end{table}

However, for the category queries (Table \ref{tab:ndcg_st_4}), we see that $p_1=1.0$ provides the worst results, while $p_1=0.4$ performs better. This means that, for this type of queries, considering the associations of the query entities with other entities mentioned in the retrieved documents affects positively the results. However, at the same time, we see that we should not give too much emphasis to this (by giving very low value to $p_1$).
Notice that this is in correspondence to the evaluation results of the probabilistic models for the same type of queries. 
Moreover, we notice that the stochastic model for $p_1=0.4$ performs better than the best probabilistic model, providing better rankings and more relevant results in the top positions.

\subsection{Synopsis of Evaluation Results}
Based on the evaluation results, we can conclude that: 
\begin{itemize}
    \item {\em Category queries} is a special case where the number of query entities can be very large and this makes the results more susceptible to disambiguation errors of the used entity linking system. Thereby, for this type of queries one should select a model which considers the  associations between the query-entities and other entities mentioned in the returned documents, since this limits the negative impact of this problem. Finally, a stochastic model with $p_1=0.4$ outperforms the best probabilistic model (relatedness).
    
    \item For the other types of queries, a model which considers both {\em relativeness} and {\em relatedness} should be selected, while {\em timeliness} does not seem to significantly affect the rankings. For single-entity and multiple-entity {\tt AND} queries, one may select the stochastic model which seems to perform better than the probabilistic models, however for multiple-entity {\tt OR} queries where the graph is not so well connected, one may opt for a probabilistic model.   
\end{itemize}

\section{Conclusions}
\label{sec:conclusion}

We have formalized the problem of ranking archived documents for structured (SPARQL) queries on semantic layers, i.e., on RDF graphs describing metadata and annotation information about the documents. 
To cope with this problem, we have proposed two ranking models, a probabilistic one and a stochastic one, which jointly consider the following aspects: i) the {\em relativeness} of the documents to the query entities, ii) the {\em timeliness} of the documents, and iii) the temporal {\em relatedness} of other entities mentioned in the documents to the query entities. 
To evaluate our approach, and due to lack of evaluation datasets for the problem at hand, we carefully created a new ground truth dataset which we make publicly available for fostering further research on similar problems. 
The evaluation results showed that the proposed models can identify important - for the query entities - documents, achieving high NDCG and precision scores and outperforming a classic, frequency-based baseline model. The results also showed that {\em relatedness}, i.e., considering the temporal association of the query entities with other entities mentioned in the results, has a high positive impact on the ranking and can limit the negative effect caused by disambiguation errors of entity linking. 

Regarding future work, we plan to study the applicability of similar models on other types of archives, like web archives where much more noisy (and probably spam) data exists and the archived documents (web pages) contain multiple similar or identical versions. 
Another interesting direction is the study of diversity-aware ranking methods that can reflect the diversity of the retrieved documents.
We also plan to build user-friendly interfaces on top of semantic layers which will make use of the proposed ranking models and will allow end-users to easily explore the archives.

\begin{acks}
The work was partially funded by the European Commission for the ERC Advanced Grant ALEXANDRIA (No. 339233).
\end{acks}

\bibliographystyle{ACM-Reference-Format}
\bibliography{jcdl2018} 

\end{document}